\documentclass[twocolumn]{aastex63}

\usepackage{hyperref}
\usepackage{verbatim}
\graphicspath{{./}{figures/}}
\received{2021 May 14}
\revised{2021 July 16}
\accepted{X, 2021}
\submitjournal{ApJL}
\usepackage{amsmath}
\usepackage{amsfonts}
\shorttitle{IBS of PSR J1959+2048}
\shortauthors{Kandel et al.}

\begin{document}

\title{XMM-Newton Observes the Intrabinary Shock of PSR J1959+2048}

\correspondingauthor{D. Kandel}
\email{dkandel@stanford.edu}

\author[0000-0002-5402-3107]{D. Kandel}
\affil{Department  of  Physics,  Stanford  University,  Stanford,  CA, 94305, USA}

\author[0000-0001-6711-3286]{Roger W. Romani}
\affil{Department  of  Physics,  Stanford  University,  Stanford,  CA, 94305, USA}

\author[0000-0002-6389-9012]{Hongjun An}
\affil{Department of Astronomy and Space Science, Chungbuk National University, Cheongju, 28644, Republic of Korea}

\begin{abstract}
In a multi-orbit (100\,ks) {\it XMM-Newton} exposure of the original black widow pulsar, PSR J1959+2048, we measure the strong orbital modulation caused by intrabinary shock (IBS) emission. The IBS light curve peak appears asymmetric, which we attribute to sweep-back effects in the companion wind. We also see evidence for an X-ray eclipse by the companion and its wind. Together with the IBS fit, this supports an edge-on $i\sim 90^\circ$ view of the system and a modest $\sim 1.8M_\odot$ mass for the recycled pulsar. Our IBS fit parameters imply a wind flux that, if persistent, would evaporated the companion within a few Gyr.  
\end{abstract}


\section{Introduction} \label{sec:intro}
The millisecond pulsar PSR J1959+2048 (hereafter J1959) was discovered at 1.4\,GHz Arecibo by \cite{fruchter1988millisecond} with a $P_S = 1.6\,$ms spin period and a $P_B=$9.16\,hr orbit with a low mass companion. The \cite{shklovskii1970possible}-corrected spindown luminosity is ${\dot E}=9.2 \times 10^{34}I_{45}\,{\rm erg\, s^{-1}}$, for a neutron star (NS) moment of inertia $I=10^{45}I_{45}\,{\rm g\,cm^2}$ and a  distance of $2\,$kpc. J1959's optical counterpart was detected by \cite{fruchter1988optical} and \cite{van1988optical} and orbital brightness modulation showed it to be tidally locked, being irradiated and evaporated by the pulsar. This evaporation inspired the moniker {\it black widow}, and it is the archetype of the companion-evaporating ``spider'' pulsars. 

Subsequent analysis of photometric data from the William Herschel Telescope (WHT) and Hubble Space Telescope (HST) by \cite{reynolds2007light} implied a binary inclination $i= 65^\circ\pm 2^\circ$, for a simple direct heating model. With this inclination, the radial velocity study of \cite{van2011evidence} implies a companion center-of-mass (CoM) velocity $K_{\rm CoM}=353\pm 4\,$km\,s$^{-1}$, giving a NS mass of $2.4\pm0.12\,M_\odot$. This large value is of substantial significance for NS Equation of State (EoS) studies. 

X-ray studies of J1959 have been carried out using ROSAT \citep{kulkarni1992x}; Chandra \citep{stappers2003x, huang2012x} and {\it XMM-Newton} \citep{huang2007xmm}, and show that the X-ray emission of J1959 is mostly non-thermal, best described by a power-law spectrum. The {\it CXO} observation resolves some of the emission into a Pulsar Wind Nebula (PWN) lying inside an H$\alpha$ bow shock. However, the compact X-ray emission presents orbital modulation, with a double-peaked lightcurve, which can be interpreted as Doppler-beamed synchrotron radiation from an intrabinary shock (IBS) between the relativistic pulsar wind and a massive wind driven from the companion \citep{Kandel_2019}. Fits with this model implied an inclination $i\sim 75.8\pm 5.9^\circ$, in some tension with the value determined from the optical light curve.

In this paper, we report on a new {\it XMM-Newton} observation of J1959. Combined with archival data sets, the improved light curve offers some potentially important clues about the binary system. In Section \ref{obs}, we briefly describe the observation and data reduction process. The spectrum and light curve analysis are presented in Section \ref{spec} and \ref{lc}, respectively. We discuss the physical implications of the results in Section \ref{Concl}. 

\section{Observations}\label{obs}
An {\it XMM-Newton} observation of J1959 in imaging mode with the medium filter was performed on May 15-16 2020 for continuous exposure of 100 ks (ObsID 0860460101). The observation was analyzed using the XMM-Newton Science Analysis System ({\tt SAS}) \citep{jansen2001xmm}. The EPIC data were processed using the {\tt epproc} and the {\tt emproc} tools, and standard pipeline processing was performed to screen particle flaring, and the events were barycentered using the {\tt barycen} tool. After screening the total live time for MOS1, MOS2 and PN were 96.9 ks, 98.6 ks and 93.4 ks, respectively. For timing and spectral analysis, a 30$''$-radius circular source extraction region centered at the position of the pulsar was used. Note that this includes extended emission from the PWN, unresolved to {\it XMM-Newton}. For background, a source-free region of approximately three times the source aperture was chosen. 

To supplement our analysis, we also use a 160\,ks archival Chandra Observation (ObsID 9088) and 31.5\,ks archival {\it XMM-Newton} observation (ObsID 0204910201). That {\it XMM} observation used PN timing mode, which results in high background, so only the MOS data are useful in our study. Unfortunately, that observation did not completely cover the orbit, missing the phase between the two peaks. Still, these data are helpful in our light curve and spectral analyses.

\section{Spectral Analysis}\label{spec}
We used the SAS tool {\tt evselect} to extract spectra of the source and the background. Response files were constructed using the SAS tools {\tt rmfgen} and {\tt arfgen}. The extracted spectra were binned with at least 25 source counts per bin. The final background-subtracted spectral modeling was performed with {\tt XSPEC} in the energy range $0.3-10.0$ keV. 

Past X-ray analysis \citep{Kandel_2019} suggested orbital variability of the spectrum of J1959, with spectrum at phases corresponding to the two IBS peaks being particularly hard, $\Gamma\sim 1$. To study this variability, we extract spectra for three different orbital phase regions: i) P1 (first peak, phase $\phi_B=0-0.25$, with $\phi_B=0$ at the ascending node of the pulsar), ii) P2 (second peak, $\phi_B=0.25-0.50$) and iii) Off-peak ($\phi_B=0.50-1.0$). At all phases (except for possibly a narrow window at $\phi_B \sim 0.25$ if there is companion eclipse), we expect a phase-independent thermal contribution from the NS surface, plus non-thermal flux from the pulsar magnetosphere and the {\it CXO}-resolved PWN. Since the IBS emission is negligible at the off-peak region, we model the off-peak spectrum as a sum of blackbody and power-law components, ${\tt BB+PL_0}$. 

Our new {\it XMM} observation spanned three complete binary orbits. During the second orbit, the flux in all cameras appears to increase substantially in the phase range $0.5-0.7$, reaching as high as $2.1\times 10^{-13}\,$erg/cm$^2$/s; over $3\times$ the quiescent level. Since other black widow systems (e.g. \citealt{romani2015spectroscopic}) are seen to have strong non-thermal flares at various orbital phases, we excise this interval in computing the typical off-peak spectrum. The best-fit for the power law in this flaring interval has $\Gamma=1.18\pm0.26$; this is nominally harder than the rest of the off-peak interval, although with limited counts the difference is not significant ($1.1\sigma$).

At P1 and P2, additional power-law emission from the IBS contributes, thus the model is taken to be ${\tt BB+PL_0+PL_i}$, with $i=1, 2$, respectively for P1 and P2. We simultaneously fit all three phase intervals (with ${\tt BB}$ and ${\tt PL_0}$ parameters, and a global $N_{\rm H}$, the same for all three intervals) to the MOS1, MOS2, and PN data. 

\begin{deluxetable}{lccc}
\tabletypesize{\footnotesize}
\tablecaption{Phase resolved Spectroscopy}
\tablehead{
\colhead{Parameter} & \colhead{P1}  & \colhead{P2} & \colhead{Off-peak}}
\startdata
      $N_{\rm H}$ ($10^{21}$cm$^{-2})$& & &$1.67\pm 0.47$ \\
     $T_{\rm{BB}}$(keV)& &  &$0.19\pm 0.02$\\
     $F_{\rm{BB}}(10^{-14}$erg/cm$^2$/s)& & &$2.03\pm 0.43$\\
     $F^\dagger(10^{-14}$erg/cm$^2$/s)& $4.37\pm 0.66$ & $5.28\pm 0.69$& $4.23\pm 0.47$\\
     $\Gamma$ &$1.28\pm 0.22$& $1.27\pm 0.18$ &$1.56\pm 0.21$ \\
    $\Gamma^*$ & $1.26\pm 0.13$ & $\Gamma_{\rm{P1}}$ & $1.53\pm 0.17$\\
\enddata
 \tablenotetext{\dagger}{Unabsorbed flux in $0.3-10.0$ keV range. The fluxes for P1 and P2 are the excess IBS flux, beyond the constant off-peak $P_0$ and blackbody fluxes.}
 \tablenotetext{^*}{Assuming fixed $N_H=1.67\times 10^{21}$cm$^{-2}$, $T_\mathrm{BB}=0.19$ keV, and $\Gamma_{\rm{P1}}=\Gamma_{\rm{P2}}$.}
\label{table:spectral_fit}
\end{deluxetable}

Since $\Gamma_1$ and $\Gamma_2$ are consistent, we also made a direct test of the spectral index variation by holding the BB and absorption fixed and simply fitting the off IBS-peak and (extra) IBS-peak power laws. The fit results are shown in Table \ref{table:spectral_fit}.

\section{Lightcurve Analysis}\label{lc}

\begin{figure}[h!]
    \centering
    \includegraphics[scale=0.35]{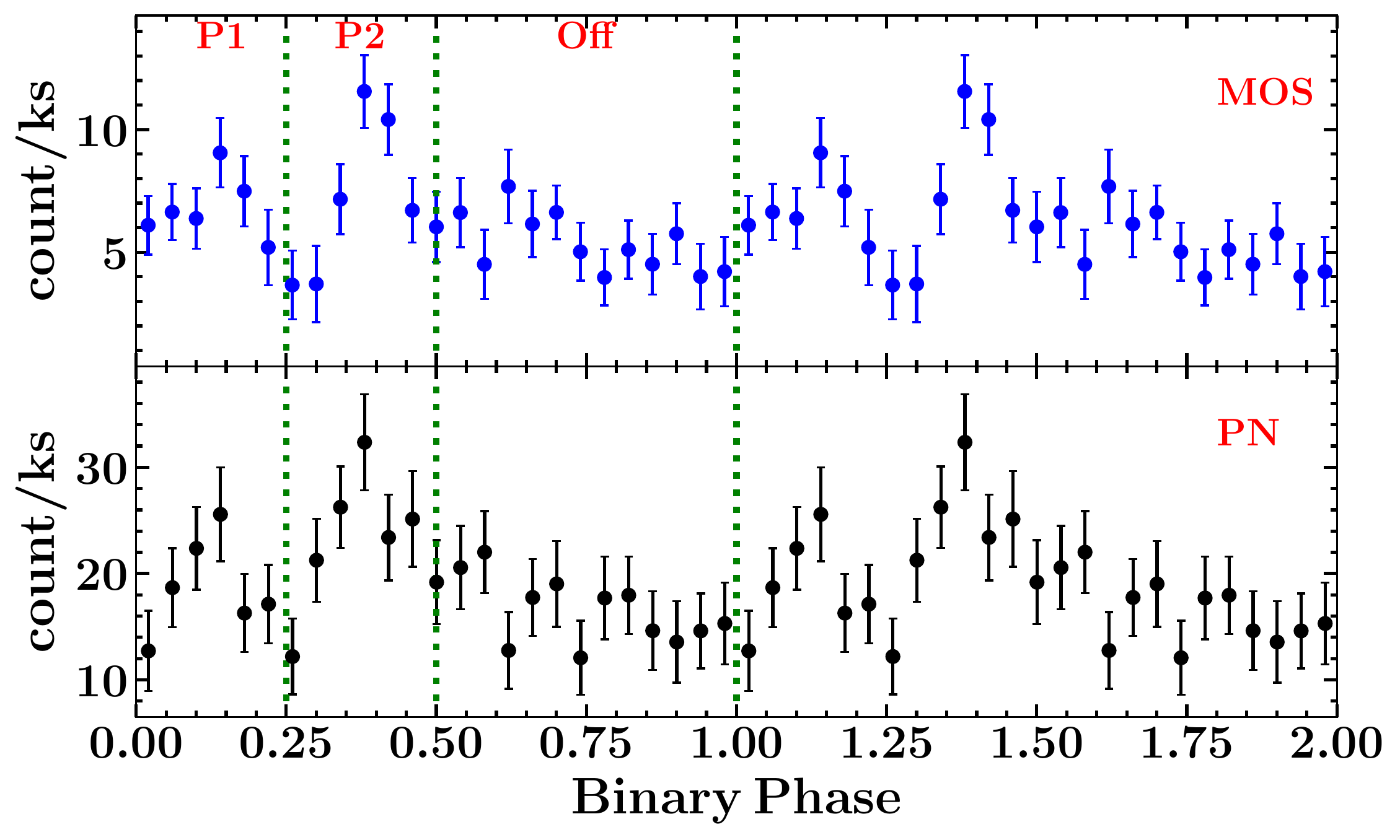}
    \caption{0.3-10\,keV LC. The upper panel shows the combined MOS1 and MOS2 LC; the lower panel shows the PN LC. Binary phase $\phi_B=0$ denotes the ascending node of the pulsar (TASC).}
    \label{fig:lc}
\end{figure}

To form our quiescent light curve (LC) we again excise the second orbit flaring interval. The resulting LC then appears fairly consistent between our new data and the archival observations. It is of course possible that lower amplitude flaring exists at other phases. Binned LCs from the MOS and PN cameras are shown in Figure \ref{fig:lc}. They show prominent double-peaked structure at phases $\sim 0.15$ and $\sim 0.45$. Between the peaks, the flux drops to the background level or below (especially in the MOS data). 

While the two bright peaks are due to IBS emission, the substantial off-peak flux is dominated by non-thermal emission from the PWN and pulsar magnetosphere and thermal emission from the NS surface. \cite{guillemot2011pulsed} showed a $4-\sigma$ detection of X-ray pulsation from J1959 and estimated that $\approx 30\%$ of the total {\it XMM} X-ray flux is pulsed. We have measured the {\it CXO} emission in the {\it XMM} aperture, excluding a $1''$-radius region around the point source. Fitting a PL spectrum (with $N_H$ fixed at the global value of \S3) we find $\Gamma_{PWN}= 1.81\pm 0.23$ and $F_{PWN} = 2.65\pm 0.38 \times 10^{-14} {\rm erg/cm^2/s}$. Thus we infer that the the blackbody emission is pulsed (as expected for a hot polar cap) and that a fraction of the PL component should also be pulsed (magnetospheric) emission. A small 10-25\% fraction of the off-peak emission appears unresolved and unpulsed; this can either be a sub-arcsec PWN component or unpulsed point source emission.

The LC also shows evidence of peak asymmetry, a feature that was hinted in \cite{Kandel_2019}, suggesting IBS sweepback distortion due to the finite speed of the companion baryonic wind; this tends to enhance and delay the second peak as seen on our LCs. Such asymmetry is characterized by the ratio between the companion's wind speed $v_{\rm W}$ and its orbital velocity $v_{\rm orb}$,  $f_{\rm v} = v_{\rm W}/v_{\rm orb}$, with lower $f_{\rm v}$ resulting in larger LC asymmetry. Accordingly, we perform a Markov Chain Monte Carlo fit to the combined LC with the model described in \cite{Kandel_2019}. For the IBS magnetic field strength, we compute the light cylinder magnetic field and assume a $1/r$ decrease (toroidal field structure) from light cylinder at $r_{L}=cP_s/2\pi$ to the termination shock, giving $B_0\sim 20 I_{45}^{1/2}$\,G for magnetic field at the nose of the IBS. We fit the shock injection power $\dot{E}_{e\pm}$, the phase-independent flux $F_0$ and the IBS parameters $\beta, f_{\rm v}$ (Table \ref{table:lc_fit}). Note that the inclination is again larger than indicated by past optical fits, although inclusion of $f_{\rm v}$ allows substantial uncertainty.

Recently, evidence has been presented \citep{Clark:2021} for an eclipse of the pulsed $\gamma-$ray emission in J1959. With a small $\sim 0.1R_\odot$ companion, this requires a binary inclination $i \approx 90^\circ$, far from the result of past optical LC modeling of this object. New optical modeling, likely including the important effects of surface heat transport and gravitational darkening \citep{kandel2020atmospheric, voisin2020model, romani2021psr} is needed to reconcile this tension. Such modeling will alter the heating pattern, but naively scaling the direct heating result of \citet{van2011evidence} one would then expect a neutron star mass $\approx 2.4M_\odot {\rm sin^3}(65^\circ) \approx 1.8M_\odot$. With an orbit close to edge-on, we also expect the point source X-ray flux (thermal and magnetospheric) to be eclipsed near $\phi_B \sim 0.25$. In addition, the IBS emission will be eclipsed. The X-ray eclipse can in fact be wider than the $\gamma$-ray event since the baryonic evaporative wind can produce significant absorption beyond the radius of the companion surface. As noted above, our data do show a flux deficit near phase $0.25$.

A simple companion photosphere eclipse of the point source would produce an X-ray eclipse width comparable to that of the $\gamma$-rays, i.e. $\Delta \phi_B \approx 0.01$. However, eclipsing X-ray flux from the extended IBS apex will produce a shallow broader modulation. More importantly, if the companion wind absorbs pulsar and IBS flux, this gives an even broader eclipse. This wind is swept back, so if the absorption takes place over a distance comparable to that of the termination shock, the absorption profile will be delayed from $\phi=0.25$ and be asymmetric. We lack the eclipse signal-to-noise ratio for a true fit, but as a first approximation to the eclipse asymmetry, we can adopt the wind column density pattern computed in \citet{an2020orbital}, assuming an equatorial wind, for the azimuthal distribution around the companion. We simplify to an asymmetric eclipser with a surface at a fixed column density in this wind pattern, so the eclipsing surface is scaled by a single parameter $r_\textrm{ecl}$ (here the minimum radius, at the azimuth of minimum column density). This surface eclipses the point source flux and the IBS. 

\begin{figure}
    \centering
    \includegraphics[scale=0.39]{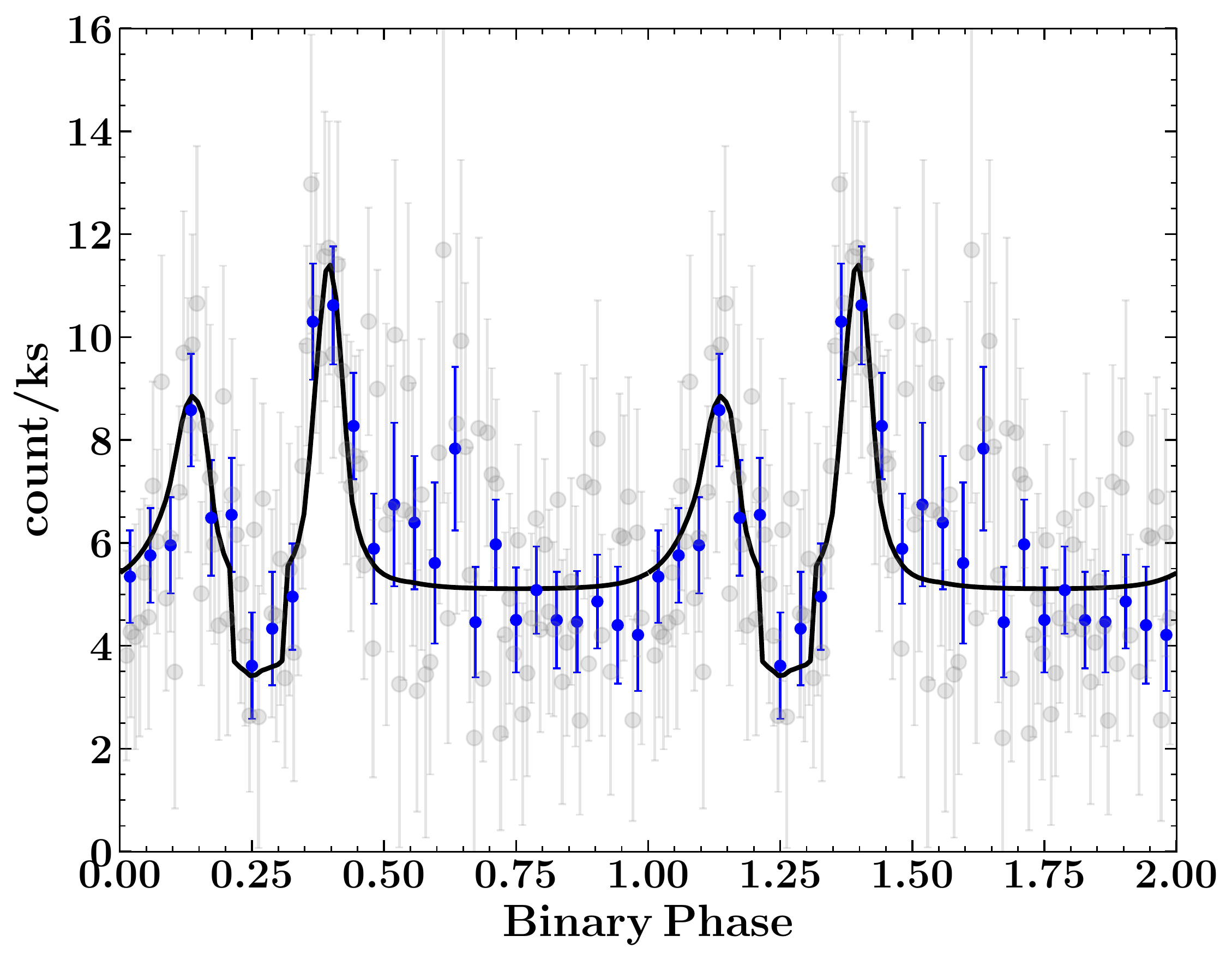}
    \caption{MOS+PN+CXO combined LC, together with the best fit model. The solid points are the combined LC data in 30 bins; the faint grey points show the bins used for fitting. The bin fluxes are converted to MOS count rate.}
    \label{fig:lcfit}
\end{figure}

\begin{deluxetable}{lcc}
\tabletypesize{\footnotesize}
\tablecaption{Lightcurve Fit Results}
\tablehead{
\colhead{Parameter} & \colhead{Free $i$, no eclipse}& \colhead{$i=90^\circ$, eclipse}}
\startdata
      $i (^\circ)$ & $78.0^{+7.7}_{-9.6}$&--\\
      $r_{\textrm{ecl}} (a)$ & -- &$0.32^{+0.19}_{-0.10}$ \\
      $\beta$ & $0.08^{+0.02}_{-0.01}$ & $0.06\pm 0.01$\\
      $f_{\rm v}$ & $14.4^{+8.3}_{-4.6}$ & $10.8^{+6.0}_{-3.5}$\\
        $\dot{E}_{e\pm}^\dagger\,(10^{34}{\rm erg\,s}^{-1})$ & $7.0\pm 0.1$& $8.4\pm 0.1$\\
      $F_0(10^{-14}{\rm erg\,cm}^{-2}\,{\rm s}^{-1})$ & $6.0\pm 0.2$ & $6.08\pm 0.22$\\ 
      $\chi^2/{\rm DoF}$ & 108/92 & 97/92\\
\enddata
 \tablenotetext{\dagger}{Assuming pulsar distance of 2\,kpc.}
 \label{table:lc_fit}
\end{deluxetable}
Fixing $i=90^\circ$, and fitting $r_{\textrm{ecl}}$ and the other IBS parameters, gives values shown in Table \ref{table:lc_fit} and the LC shown in Figure \ref{fig:lcfit}. The inferred eclipsing wind is extended with $r_{\textrm{ecl}}\approx 0.32a$, at the smallest, extending to nearly $a$, the orbit semi-major axis, in the swept-back direction. The dip centered at phase $>0.25$ is clearly captured by the model; so is the peak asymmetry. The decrease in $\chi^2$ by adding an eclipse is $\sim 10$ (see Table \ref{table:lc_fit}), thus the eclipse model is statistically preferred: the non-eclipse model's relative likelihood is about 0.5\% according to the Akaike Information Criterion.

\section{Discussion and Conclusions}\label{Concl}

The spectral results are in good accord with prior fits. The off-peak power law emission includes flux from the inner nebula and our $\Gamma=1.53 \pm 0.17$ is consistent with the value for the inner nebula, resolved with {\it CXO}, as found here and in \citet{huang2012x}. The peak emission appears harder than off-peak, but the significance remains marginal. Even holding the thermal emission and $N_{\rm H}$ fixed, the mean index of the peak region, $\Gamma^\ast=1.26$, differs by only $1.3\sigma$ from the off-peak power law, or $2.1\sigma$ from the extended {\it CXO} PWN spectral index. We are however able to separate a thermal component in the off-peak phase, which is consistent with a heated polar cap of radius 0.4\,km and temperature $2.2 \times 10^6$\,K; this flux should be pulsed and may be eclipsed by the companion. 

The fixed-$i$ fit to the IBS peak, Table \ref{table:lc_fit}, provides substantially improved constraints on the shock properties. First, ${\dot E}_{e\pm}$, the isotropic equivalent power required to explain the IBS luminosity at $d=2$\,kpc is quite close to the spindown power. Even with $I_{45}>1$ or an expected equatorial concentration of the pulsar wind, the efficiency will be high; we can see this agreement as a crude confirmation of the source distance. The wind speed parameter $f_{\rm v}$ is now measured with some confidence. Combined with the model parameter $\beta\equiv \dot{M}_{\rm W}v_{\rm W}c/\dot{E}$, which describes the momentum ratio of the companion and pulsar winds, and $\dot{E}$ from the pulsar spindown we can estimate the companion mass loss rate as
\begin{equation}\label{eq:mass_loss}
    \dot{M}_{\rm W} = \frac{\beta\dot{E}}{f_{\rm v}cv_{\rm orb}}~.
\end{equation}
With a companion orbital velocity $350\,$km\,s$^{-1}$, our best-fit results gives a mass-loss rate of $\dot{M}_{\rm W}\approx 8_{-3}^{+4}\times 10^{-12}I_{45}M_\odot\,{\rm yr}^{-1}$. This mass loss rate is at least $5\times$ higher than the $\sim 10^{-12}M_\odot\,{\rm yr}^{-1}$ inferred from radio eclipse by \citet{2020MNRAS.494.2948}. The difference may not be too surprising since our method relies on the global structure of the wind shock, which is directly sensitive to the mass flux, rather than the less direct inference from the radio eclipse provided by the ionized component. This instantaneous mass loss rate is interesting as, even at the 1--$\sigma$ lower limit, complete evaporation would take $5\times 10^9$\,yr, assuming a companion mass of $\sim 0.024M_\odot$. Of course, the mass loss may decrease as the system evolves, but the present few-Gyr timescale implies that while we are observing J1959 in a long-lived phase, complete evaporation to an isolated millisecond pulsar could be possible in the spindown lifetime.

\begin{figure}
    \centering
    \vskip -1.0truecm
    \includegraphics[scale=0.4]{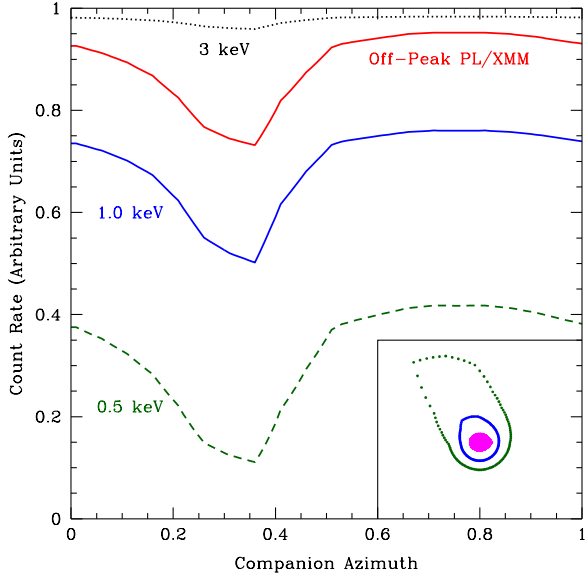}
    \vskip -2.0truecm
    \caption{X-ray absorption in the swept-back companion wind will be asymmetric and delayed from $\phi_B=0.25$. Here we show a toy model for the absorption in the companion wind, as a function of companion azimuth, for three X-ray energies and for the total count rate of the off-peak power-law emission, folded through the {\it XMM} PN response. The boxed sketch on the lower right shows the companion photosphere (PSR to the right, not shown) and dotted lines for the absorption wind effective radius at two energies. Our toy model simply assumes occultation by a companion of the appropriate absorption wind radius. A detailed computation of the eclipse would include energy-dependent absorption of the PSR and IBS emission for all sightlines through the wind at each orbital phase.}
    \label{fig:Xabs}
\end{figure}

This X-ray eclipse provides an opportunity to constrain the companion wind structure and mass flux. At high energies (hard X-ray to $\gamma$-ray), only the dense companion photosphere at radius $r_\ast$ will eclipse. At progressively lower X-ray energies the extended, swept-back companion wind will provide more absorption, with increasing optical depth and increasing distortion (Figure \ref{fig:Xabs}). We defer such detailed computation to a future publication; next-generation  X-ray sensitivities will likely be required to take advantage of the energy-dependent eclipse to probe the wind structure. We do, however, note a basic inference from the apparently extended eclipse. An eclipse with characteristic scale $a \approx 2 \times 10^{11}{\rm cm}$ will have a column density $\rho \sim {\dot M} /(2\pi r_\ast f_{\rm v} v_{\rm orb})$ for a near-equatorial wind ($r_\ast\rightarrow 0.3a$ for an approximately spherical outflow). This corresponds to a hydrogen column density $\sim 7 \times 10^{19} I_{45}/(f_{\rm v}/10)^2 {\rm cm^{-2}}$. Since we typically infer an interstellar column density $\sim 4 \times 10^{21}{\rm cm^{-2}}$ for an absorption optical depth $\tau \sim 1$ at 1\,keV, the wind seems $\sim30/I_{45}\times$ too thin to give a strong absorption eclipse. More detailed sums, e.g. including pile-up at the termination shock, may address this. But it also seems important to note that spectroscopic studies of J1959 \citep{van2011evidence} and other black widows \citep{romani20142fgl, romani2015spectroscopic} companions infer metallicities substantially higher than Solar. This reduces the required column for $E>$\,keV absorption, providing appropriate columns for $[Z/H] > 10$. In the {\it Athena}/{\it Lynx} future, one might use detailed-energy dependent eclipse curves to probe the wind composition, in addition to its density structure.

\acknowledgements
We thank the referee, whose comments helped us improve the paper.
This work was supported in part by NASA grants 80NSSC17K0024 and 80NSSC21K0896 .

\bibliographystyle{aasjournal}
\bibliography{mainpaper}
\end{document}